\documentclass{nature}
\usepackage{graphicx}

\makeatletter
\let\saved@includegraphics\includegraphics
\AtBeginDocument{\let\includegraphics\saved@includegraphics}
\renewenvironment*{figure}{\@float{figure}}{\end@float}
\makeatother

\usepackage{amsmath,amsfonts,amsthm,amssymb}
\usepackage{xcolor}
\usepackage[colorlinks=true,linkcolor=blue,citecolor=blue,urlcolor=blue]{hyperref}
\usepackage[capitalise]{cleveref}
\usepackage[utf8]{inputenc}
\usepackage[english]{babel}
\addto\captionsenglish{}
\usepackage{physics}
\usepackage{ccaption}

\newcommand{\ill}{ Institut Laue-Langevin, 71 Avenue des Martyrs, 38042 Grenoble, France}
\newcommand{\liphy}{ Université Grenoble Alpes, CNRS, LIPhy, Grenoble, France}
\newcommand{\ens}{ Laboratoire de Physique de l’Ecole Normale Supérieure, CNRS, Université PSL, Sorbonne Université, Paris, Paris, France}

\usepackage[normalem]{ulem}

\title{Fluids at an electrostatically active surface: Optimum in interfacial friction and electrohydrodynamic drag}

\author{Cecilia Herrero$^{1\,2\,\bigstar}$, Lyd\'eric Bocquet$^{2}$ \& Benoit Coasne$^{1\,3\,\bigstar}$}

\begin{document}
	
	\maketitle
	
	\begin{affiliations}
		\item{\ill}
		\item{\ens}
		\item{\liphy}	
		\item[$^\bigstar$] e-mail: cecilia.herrero@ens.fr, \,
benoit.coasne@univ-grenoble-alpes.fr
	\end{affiliations}	
	
	\begin{abstract} While fluids near a solid surface are at the core of applications in energy storage/conversion, electrochemistry/electrowetting and adsorption/catalysis, their nanoscale behavior remains only partially deciphered. Beyond conventional effects (e.g. adsorption/reaction, interfacial transport, phase transition shifts), recent experimental and theoretical studies on metallic surfaces have unraveled exotic peculiarities such as complex electrostatic screening, unexpected wetting transition, and interfacial quantum friction. These novel features require developing and embarking new tools to tackle the coupling between charge relaxation in the metal and molecular behavior in the vicinal fluid. Here, using the concept of Virtual Thomas-Fermi fluids, we employ a molecular simulation approach to investigate interfacial transport of fluid molecules and metal charge carriers at their interface -- including the underlying electrostatically-driven dynamic friction and the coupling between charge current/hydrodynamic flow (the so-called \textit{electrohydrodynamic drag}). While conventional numerical techniques consider either insulating materials or metallic materials described as polarizable, non-conducting media, our atom-scale strategy provides an effective yet realistic description of the solid excitation spectrum -- including charge relaxation modes and conductivity. By applying this approach to water near metallic surfaces of various electrostatic screening lengths, we unveil a non-monotonous dependence of the fluid/solid friction on the metallicity with a maximum occurring as the charge dynamic structure factors of the solid and fluid strongly overlap. Moreover, we report a direct observation of the \textit{electrohydrodynamic drag} which arises from the momentum transfer between the solid and liquid through dynamic electrostatic interactions and the underlying interfacial friction. 
	\end{abstract}
	\maketitle
	


\noindent Fluid/solid interfaces as involved in nanofluidic devices or in porous materials are the host of a wealth of microscopic mechanisms including molecular adsorption, chemical reaction, interfacial friction, local segregation, etc.\cite{bocquet2010nanofluidics,huber2015soft,kavokine2021fluids,schlaich2024theory} While many of these molecular phenomena can be harnessed in nanotechnologies for energy storage, catalysis, lubrication, depollution, etc., the underlying microscopic processes also combine at the nanoscale to give rise to novel effects which cannot be rationalized using currently available formalisms. Of particular relevance to the applications above, several reports have recently unveiled the role of the physical coupling between the behavior of the interfacial fluid and charge relaxation within the solid material. Such effects, which go beyond well-documented effects involving interfacial slippage and surface corrugation,\cite{bocquet1994hydrodynamic,falk2010molecular,herrero2020fast,oga2021theoretical,lizee2024anomalous} molecular intermittence and memory effects,\cite{levitz2005random,bousige2021bridging} nanoscale wetting at defective or heterogeneous surfaces,\cite{coasne2008adsorption,rotenberg2010coarse,thiemann2024defects} and adsorption/desorption mechanisms,\cite{serva2021confining,mangaud2022chemisorbed} point to the need to account for the strong interplay that occur between the solid and fluid phases at their interface.  As far as thermodynamic aspects are concerned, the surface metallic/insulating nature -- as described through the screening length $\ell_{\rm TF}$ which characterizes the distance over which charge relaxation occurs -- was found to drastically affect the wetting of an interfacial ionic liquid and the confinement-induced shift in its freezing point.\cite{comtet2017nanoscale, kaiser2017electrostatic,schlaich2022electronic} As for dynamical aspects, such charge relaxation within metallic systems was also found to lead to a \textit{quantum friction} which stems from the overlap of the excitation spectrum of the charge carriers in the solid and the charge dynamic structure factor of the molecules in the interfacial fluid.\cite{kavokine2022fluctuation} Such a non-conventional phenomenon and the underlying dynamic electrostatic coupling in fluids near metallic surfaces was predicted to lead to unprecedented transport mechanisms. This includes electrohydrodynamic drag effects which correspond to the cross-talk between a fluid molecular flow and a charge current within the metal at their interface.\cite{coquinot2024hydroelectric} Other related interfacial mechanisms include the fact the molecular flow in an adsorbed fluid can set in motion another adsorbed fluid separated by a metallic material of a nanoscale height.\cite{coquinot2025momentum}

\noindent This fascinating impact of surface metallicity on the behavior of interfacial fluids has led to a new momentum in the field. While theoretical approaches on nanoconfined fluids have mostly focused until recently on perfect metallic or insulating surfaces (except for a few contributions accounting for electrostatic screening\cite{newns1969fermi,inkson1973many,kornyshev1977image,netz1999debye,kornyshev2014differential,kaiser2017electrostatic}), several molecular simulation strategies are now available to mimic charge relaxation within the metallic material to induce a physical screening length as encountered in real materials. Such numerical approaches\cite{scalfi2020semiclassical,scalfi2021microscopic,bui2023classical} involve a solid surface made up of an underlying atomic structure that bears localized polarizable charges. On the one hand, a possible approach is to mimic electrostatic relaxation within the material by associating around each solid atom a Gaussian charge distribution that relaxes in response to any change in the fluid molecular configuration.\cite{scalfi2020semiclassical,scalfi2021microscopic} On the other hand, charge relaxation within the solid can also be mimicked using a Drude oscillator model in which an electrostatic charge is linked to each solid atom through a spring.\cite{bui2023classical} While these strategies mimic in an effective fashion the physics of electrostatic screening in the solid and its impact on the fluid at its surface, the underlying description of the charge carriers implies that the metal is treated as a polarizable but non-conducting medium. As a result, of particular relevance to the topic of quantum friction and related interfacial effects, these approaches fail to capture the coupling between fluid transport (molecular flow) and charge transport (current) at fluid/metal interfaces. In particular, charge transport within the solid itself -- which is an essential ingredient to unravel the molecular mechanisms leading to hydroelectric energy conversion by harnessing the dynamic liquid/solid interplay at their interface\cite{coquinot2023quantum} -- requires an approach that mimics both electrostatic screening and charge transport within the metal. 

\noindent In this context, we note that such intertwined dynamic phenomena can be captured in an effective yet realistic manner using classical molecular simulations involving Virtual Thomas-Fermi fluids.\cite{schlaich2022electronic} As described below, with this strategy, which allows treating metallic surfaces of any geometry and/or screening length, electrostatic screening and charge transport in the confining metal are mimicked by considering light mobile charged particles that relax in the presence of the interfacial fluid. Here, by invoking the concept of Virtual Thomas-Fermi fluids, we implement an extended molecular simulation approach which allows us to unravel the microscopic mechanisms that govern the coupled transport of fluid water molecules and metal charge carriers at their interface.  Thanks to the use of this classical method that describes in an approximate fashion but physical fashion the whole solid microscopic behavior (including its conductivity), we investigate the impact of the electrostatic screening length of the metallic surface on the fluid/solid friction. Using the conventional Green-Kubo formalism involving the time correlations of the solid/fluid force at their interface, we unveil an optimum in the interfacial friction upon varying the solid metallicity -- from a perfect insulator towards a perfect metal. By invoking the charge dynamic structure factors $S_{\rm cc}(k,\omega)$ of the fluid and solid set in contact, we show that this optimum occurs as their energy/wavevector spectra strongly overlap. In more detail, while the spectrum of vicinal water is only weakly modified upon tuning the solid metallicity, that of the solid shifts in a continuous fashion so that a strong coupling in the fluid and solid microscopic dynamics can be achieved when the overlap is maximum. Then, we extend our molecular simulation strategy to non-equilibrium transport situations to investigate how such dynamic \textit{quantum friction} between the fluid and solid leads to coupled molecular and charge flows. Finally, despite the use of a classical approach, we observe a direct correlation between such molecular and charge transport with the the emergence of the so-called \textit{electrohydrodynamic drag} and its reciprocal effect which has been coined as \textit{quantum osmosis}.\cite{coquinot2024hydroelectric}.

\begin{figure}[h!]
    \centering
    \includegraphics[width=0.8\textwidth]{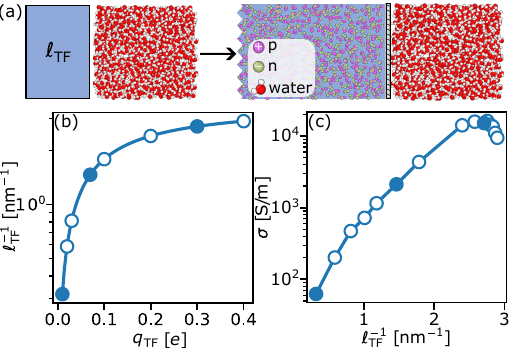}
    \caption{{\bf Effective Coulomb screening at a solid/fluid interface using a Virtual Thomas-Fermi fluid.}
    \textbf{a}, Schematic illustration of the electrostatic screening as modeled using the concept of Virtual Thomas-Fermi fluids. The electrostatic screening induced in the vicinity of a metallic interface is characterized by a screening length $\ell_{\rm TF}$ which can take values from 0 (perfect metal) to any non-zero value (imperfect metal). With the Virtual Thomas-Fermi method, the solid is described as a set of positive and negative charges having a charge $q_{\rm TF}$. Here, we model water at the surface of the metallic system. The two systems [liquid water as the physical fluid and the Virtual Thomas-Fermi fluid to mimic the metallic surface]  are separated by a reflective wall to prevent mixing.  (b) Inverse screening length of the solid $\ell_{\rm TF}^{-1}$ as a function of the charge $q_{\rm TF}$ of the particles in the Virtual Thomas-Fermi fluid. The system can be tuned continuously from a perfect insulator ($\ell_{\rm TF}^{-1} = 0$) towards the perfect metal limit ($\ell_{\rm TF}^{-1} \to \infty$).
    (c) Solid electrical conductivity $\sigma$ as described using the Virtual Thomas-Fermi method as a function of the inverse screening length $\ell_{\rm TF}^{-1}$.}
    \label{fig:fig1}
\end{figure}

\noindent {\bf Virtual Thomas-Fermi fluids and effective Coulomb screening}\\
\noindent As illustrated in Fig.~\ref{fig:fig1}(a), we employ the Virtual Thomas-Fermi technique to investigate the impact of surface metallicity on the dynamics of water in the vicinity of a solid surface. Water was selected as an ideal candidate to investigate the dynamic correlations between the fluid molecular flow and charge current within the metal at their interface. Indeed, the dynamic spectrum of liquid water displays typical excitations $\hbar \omega$ in the Terahertz (THz) frequency range which are expected to couple with the electronic spectrum of typical metals. This makes water an ideal candidate to unravel the molecular mechanisms underlying transport at fluid/metal interfaces. The Virtual Thomas-Fermi method is an original method to simulate in an effective fashion electrostatic screening induced within a metal as charge carriers relax.\cite{schlaich2022electronic} With this classical method, the charges within the metal are modeled by considering a set of positive and negative light charges which relax on a much shorter timescale than the fluid at the metal interface (the effective screening length is directly related to the electrostatic charge of these particles like in the concept of Debye length for liquid electrolytes). While adopting a hydrodynamic (i.e. classical) description of charge relaxation in metals is unconventional and necessarily approximate, we note that it was formally demonstrated that Boltzmann transport equation can be recovered within such frameworks.\cite{pitarke2006theory} Through simple implementation in a Monte Carlo or Molecular Dynamics simulation strategy, the Virtual Thomas-Fermi approach was found to successfully reproduce the capacitive behavior of metallic interfaces while also capturing the experimentally observed impact of surface metallicity on the freezing of a charged liquid.\cite{comtet2017nanoscale} This method also yields unprecedented predictions such as a wetting transition as the solid's properties are tuned from an insulator towards a metal.\cite{schlaich2022electronic}

\noindent In our Molecular Dynamics strategy, the charges in the Virtual Thomas-Fermi fluids are allowed to relax fast by using the following mass: $m_+ \sim 0.1$~$m$ for cations and $m_- \sim 0.005~m$ for anions (where $m$ is the mass of the H atom). In practice, as shown in Fig. 1(a), the effective Virtual Thomas-Fermi method consists of placing the dipolar or charged fluid (here, water) in the vicinity of the metallic solid. To prevent mixing between the solid and fluid phase, a reflective wall is introduced at their interface. The domain corresponding to the metallic solid is filled with the Virtual Thomas-Fermi fluid with a density $\rho_{\rm TF}$. In its original version, the Virtual Thomas-Fermi was implemented by considering charge carriers in the metal at extremely high temperatures ($T \sim$12000~K) to ensure very fast relaxation. While this is physically acceptable (since this method relies on a virtual, i.e. effective, fluid which needs not to possess physical properties), this practical implementation leads to a non-physical temperature transfer at the metal/fluid interface which makes it impossible to investigate their coupled transport. To circumvent this caveat, as described below and in more detail in the Methods section, we extended the Virtual Thomas-Fermi strategy to room temperature conditions while carefully avoiding crystallization of the charges within the metal. Since the explicit charges in the Virtual Thomas-Fermi fluid are used without any solvent, we used a relative dielectric constant $\varepsilon_{\rm r}= 1$. All technical details are provided in the Methods section.

\noindent We address the previous limitations in the Virtual Thomas-Fermi method by modeling the solid as a set of mobile charges having charge $|q_{\rm TF}|$ and a density $\rho_{\rm TF}$ to produce a tunable screening length $\ell_{\rm TF}$. As a first-order approximation, this screening length $\ell_{\rm TF}$ can be taken as the Debye length $\ell_{\rm D}$ for the metallic region (blue area in Fig.~\ref{fig:fig1}a):
\begin{equation}
	\ell_{\rm TF} \sim \ell_{\rm D} = \sqrt{\frac{\varepsilon_0 k_{\rm B} T}{\rho_{\rm TF} q_{\rm TF}^2}},
	\label{eq:ellTF}
\end{equation}
where $\varepsilon_0$ is the dielectric constant of vacuum and $T$ is the temperature of the Virtual Thomas-Fermi fluid (which is taken equal to the temperature of the fluid phase as discussed earlier). 
This framework efficiently captures electrostatic interactions near metals while enabling the simulation of a wide range of insulating and metallic behaviors. By tuning $\ell_{\rm D}$, this model continuously transitions from perfect metal ($\ell_{\rm D} = 0$) to a perfect insulator ($\ell_{\rm D} \to \infty$). While the Debye length in Eq.~(\ref{eq:ellTF}) provides an approximate value for the screening length, we determined the exact screening length $\ell_{\rm TF}$ imposed by the Virtual Thomas-Fermi fluid from capacitance assessments. In more detail, as detailed in the Methods section, we consider a vacuum slab sandwiched between two electrodes filled with Virtual Thomas-Fermi fluids to obtain an accurate estimate of the metal screening length $\ell_{\rm TF}$. To illustrate the versatility of our approach, Fig.~\ref{fig:fig1}(b) shows the inverse screening length  $\ell_{\rm TF}^{-1}$ as a function of the simulation parameter $q_{\rm TF}$ (which we vary to tune the solid's metallicity at constant charge density $\rho_{\rm TF}$ and temperature $T$). In practice, the screening length $\ell_{\rm TF}$ can be tuned by varying $q_{\rm TF}$ from $q_{\rm TF} = 0~e$ [$\ell_{\rm TF} \to \infty$] for an insulator to $q_{\rm TF} = 0.4~e$ [$\ell_{\rm TF} \to 0.345$~nm] for a nearly perfect metal ($e$ is the elementary charge). Interestingly, experimental $\ell_{\rm TF}^{-1}$ for real materials span a broad range which is compatible with the values mimicked using the Virtual Thomas-Fermi approach: from $\ell_{\rm TF}^{-1}=0~$nm$^{-1}$ for mica to $\ell_{\rm TF}^{-1}\sim 7~$nm$^{-1}$ for platinum with intermediate values for HOPG and doped silicon.\cite{comtet2017nanoscale} While our approach cannot exactly reproduce the perfect metal limit, we note that this asymptotic situation can be accurately handled  by relying on the image charge method.\cite{iori2008including}

\noindent With the goal to further assess our ability to describe the solid metallicity using the Virtual Thomas-Fermi approach, Fig.~\ref{fig:fig1}(c) shows the solid conductivity  $\sigma$ as a function of the inverse screening length $\ell_{\rm TF}^{-1}$ (all details can be found in the Methods section). On the one hand, starting from an insulating material, the conductivity $\sigma$ first increases with $\ell_{\rm TF}^{-1}$ as the material evolves from insulating to conductive. In this regime, the conductivity increases as the conductivity is directly related to the density of charge carriers in the Virtual Thomas-Fermi fluid. On the other hand, as the screening length reaches $\ell_{\rm TF}\sim 2.7~$nm$^{-1}$, a maximum is observed in $\sigma$ as the conductivity decreases upon further increasing  $\ell_{\rm TF}$. Such a decrease in the  conductivity occurs due to strong ion pairing effects which reduce ion mobility. 



\begin{figure}[h!]
	\centering
	\includegraphics[width=0.99\textwidth]{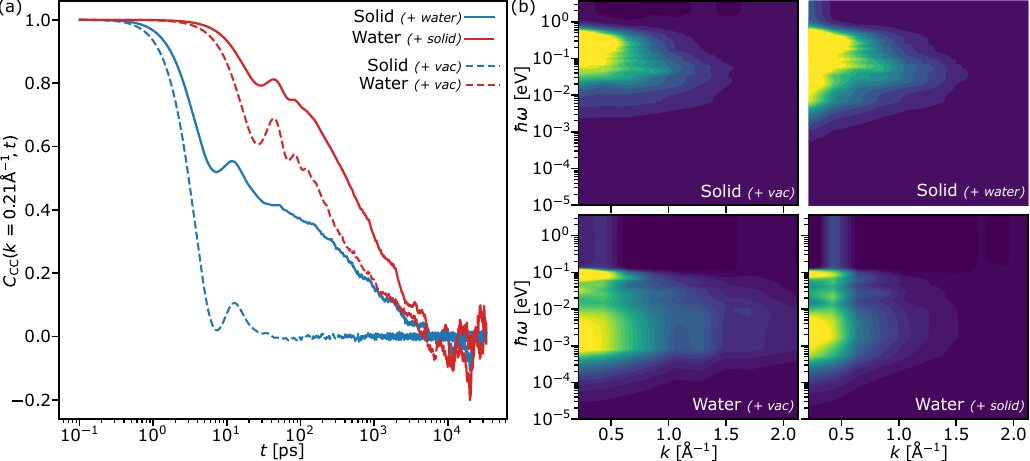}
	\caption{{\bf Surface response functions and solid/fluid coupling at their interface.}	(a) Normalized charge/charge correlation function $C_{\rm cc}(k,t)$ for a wavevector $k = 0.21~$\AA$^{-1}$ as a function of time $t$. The blue data correspond to the charge dynamic structure factor for the solid surface while the red data correspond  to the charge dynamic structure factor for the solid surface. As shown in the legend, the dashed lines correspond to the solid and water phases in contact with vacuum while the solid lines correspond to the solid and water phases in contact with each other. The longer relaxation timescales observed when the solid and water are set in contact indicate a slowdown induced by the friction/coupling between the two phases. (b) Response function ${\rm Im}[g(k,\omega)]$ which illustrates the full wavevector $k$ and energy transfer $\hbar \omega$ spectra. The results correspond to a screening length $\ell_{\rm TF}^{-1} = 2.7~$nm$^{-1}$, which is associated with the maximum conductivity shown in Fig.~\ref{fig:fig1}(c). These data highlight the  interplay between the wavevector-dependent dynamics of the interfacial fluid and the electrostatic response function of the solid surface.
	}
\label{fig:fig2}
\end{figure}

\noindent {\bf Dynamic coupling at the solid-liquid interface and  interfacial \textit{quantum} friction}\\
\noindent To investigate the electrostatic correlations at the interface between the solid charge carriers and the water fluid molecules, we first assess their microscopic dynamics through the charge/charge time correlation function $C_{\rm cc}(k,t)$ for a given wavevector $k$:
\begin{equation}
	C_{\rm cc}(\vb{k},t) = \frac{\langle n_{\rm cc}(\vb{k},0)n_{\rm cc}(\vb{k},t) \rangle}{\langle n_{\rm cc}(\vb{k},0)^2 \rangle},
\end{equation}
where $\langle \cdots \rangle$ denotes statistical ensemble average and $n_{\rm cc}(\vb{k},t) = \sum_j q_{j} \exp[i \vb{k} \cdot  \vb{x}_j]\exp[-k \abs{\delta z_j}]$ is the Fourier component $\vb{k}$ of the surface charge density $n_{\rm cc}(\textbf{r},t)$. In the last expression, the sum runs over all charges $j$ having a charge $q_j$ and $\vb{r}_j = (\vb{x}_j,z_j)$ is the charge position. $\delta z_j = z_j - z_0$ is the distance between the charge and the solid/fluid interface which is located at a position $z_0$ (the latter is defined as the center of the reflective wall that prevents mixing between the physical fluid and the Virtual Thomas-Fermi fluid). Since our system in isotropic in the $x$-$y$ plane, all results below specifically refer to the wavevector $\vb{k}=(k,0,0)$ so that $k = |\vb{k}|$.  Fig.~\ref{fig:fig2} shows the charge/charge correlation function $C_{\rm cc}(k,t)$ for the surface water film and the Virtual Thomas-Fermi fluid in the case $\ell_{\rm TF}^{-1} = 2.7~$nm$^{-1}$ [such electrostatic screening length was chosen as it corresponds to the conductivity maximum shown in Fig.~\ref{fig:fig1}(c)]. In more detail, we present the time correlation functions obtained for the following distinct configurations: ($i$) the solid in the presence of water, ($ii$) the water film in the presence of the solid, ($iii$) the solid in contact with vacuum (no water film),  and ($iv$) the water film in contact with vacuum (no Virtual Thomas-Fermi fluid). For both the solid and water film, the charge/charge correlation function $C_{\rm cc}(k,t)$ decays over significantly longer relaxation timescales when the virtual Thomas-Fermi and water film are set in contact (see comparison with the data obtained when each phase is set in contact with vacuum). Such an increase in the charge/charge relaxation time provides evidence for a dynamic coupling between the solid and fluid at their interface. In other words, as will be established below, the coupling between the microscopic relaxation phenomena within the interfacial fluid and metal gives rise to an dynamic electrostatic friction which slows down the overall fluid and solid dynamics. 

\noindent Following the seminal work by Kavokine et al.,\cite{kavokine2022fluctuation} we further identified the dynamical spectra of the charge surface density by determining the imaginary part of the so-called surface response function:
\begin{equation}
	\operatorname{Im}[g(k,\omega)] = \frac{e^2}{4 \varepsilon_0 k} \frac{\omega}{k_{\rm B}T}S_{\rm cc}(k,\omega),
\end{equation}
where $e$ is the elementary charge, $\varepsilon_0$ the dielectric constant of vacuum, and $T$ the temperature. $S_{\rm cc}(k,\omega)$ is the charge/charge dynamic structure factor which is obtained as the time Fourier transform of the charge/charge autocorrelation function per unit of interface area $\mathcal{A}$:
\begin{equation}
	S_{\rm cc}(k,\omega) = \frac{1}{\mathcal{A}}\int {\rm d}t ~ \langle n_{\rm cc}(k,0) n_{\rm cc}(k,t) \rangle \exp[{i\omega t}],
\end{equation}
where we recall that $\langle \cdots \rangle$ denotes statistical ensemble average. The response functions $\operatorname{Im}[g(k,\omega)]$ provide the complete dynamic spectrum of the solid and fluid phases as they contain the density of states of excitations corresponding to a wavevector $k$ and energy transfer $\hbar \omega$. As illustrated in Fig.~\ref{fig:fig2}(b) for the case $\ell_{\rm TF}^{-1} = 2.7~$nm$^{-1}$, both the metallic solid and water film are affected as they are set in contact. Indeed, comparison between the surface responses for the solid phase (water phase, resp.) in the absence and in the presence of the water phase (solid phase, resp.) shows that the dynamics of both phases are coupled. In more detail, while the water microscopic dynamics (right panels) is only little affected by the presence of the solid, the solid microscopic dynamic spectrum exhibits a marked slowdown when set in contact with water. In more detail, such slowdown is observed as the whole spectral response of the solid phase is shifted towards smaller energy transfers $\hbar \omega$ (or, equivalently, longer relaxation times). We emphasize that these results, which reveal an intricate coupling between the solid and fluid phases, cannot be attributed to momentum transfer via phonons at the fluid/solid interface as our model does not involve any underlying atomic structure.



\noindent The fluid/solid dynamic coupling as revealed through the analysis of the charge/charge correlation functions and associated surface responses has profound implications for interfacial transport. The intertwined dynamics directly impacts the fluid/solid friction, which quantifies the resistance to the relative motion of the two phases at their interface. In particular, the friction parameter $\lambda$ serves as an essential transport coefficient that governs the molecular flow and energy dissipation at the interface. Classically, interfacial friction is understood to arise primarily from surface roughness; as water molecules flow over a solid surface, they collide with the microscopic lattice structure of the solid so that momentum is transferred from/to the solid lattice vibrations (phonon modes). Yet, as recently established in the context of fluid films in the vicinity of metallic surfaces, an additional contribution -- coined as \textit{quantum friction} -- arises from the coupling between the dynamic charge relaxations within the metal and the molecular dynamic modes within the interfacial fluid. In this context, we employ our effective strategy based on Virtual Thomas-Fermi fluids to investigate the impact of surface metallicity on such quantum friction. Here, we emphasize that this simple yet robust approach constitutes an ideal playground to investigate such novel physical phenomena as the absence of solid atomic structure eliminates any classical contribution to interfacial friction (therefore, ensuring that the observed interfacial friction only arises from the coupling with electrostatic interactions/relaxation within the metallic phase).

\begin{figure*}
	\centering
	\includegraphics[width=0.99\textwidth]{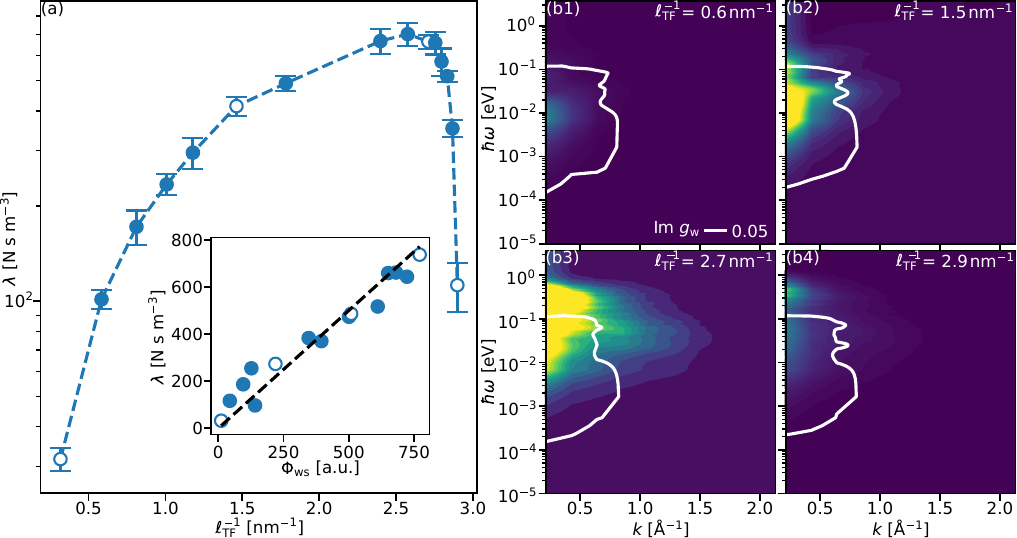}
	\caption{{\bf Interfacial friction and overlap between the solid/fluid response functions.}
	(a) Fluid/solid friction coefficient $\lambda$ as a function of the inverse screening length $\ell_{\rm TF}^{-1}$. $\lambda$ is calculated from equilibrium Molecular Dynamics using the Green-Kubo integral of the force time autocorrelation function. A maximum in friction is observed at a screening length of approximately $2.7~\mathrm{nm}^{-1}$.  The inset shows $\lambda$ as a function of the water/solid interaction spectrum, $\Phi_{\rm ws}$, which represents the summation of the imaginary parts of the water and solid surface responses over all possible $k$ and $\omega$ values. As indicated by the black dashed line, a perfect correlation is observed: $\Phi_{\rm ws} = \lambda$. (b) Solid response functions ${\rm Im}[g_{\rm s}(k,\omega)]$ as a function of the wavevector $k$ and energy transfer $\hbar \omega$ for four different inverse screening lengths $\ell_{\rm TF}^{-1}$ [the latter values are those shown as empty markers in (a)]. The white line  corresponds to the water response function $\operatorname{Im}[g_w(k,\omega)]$ taken at the following isovalue condition $\operatorname{Im}[g_w(k,\omega)] = 0.05$. As explained in the text, the overlap of these two response functions corresponds to $\Phi_{\rm ws}$ -- therefore demonstrating that the maximum friction occurs when the two spectra significantly overlap.
	}
\label{fig:fig3}
\end{figure*}

\noindent With the aim to determine the impact of electrostatic screening  on interfacial friction, we determine $\lambda$ from the well-established Green-Kubo relation \cite{bocquet1994hydrodynamic}:
\begin{equation}
	\lambda = \frac{1}{\mathcal{A}k_{\rm B}T} \int {\rm d} t \langle F(0)F(t) \rangle,
\end{equation}
where $k_{\rm B}$ is Boltzmann's constant and $F(t)$ is the force experienced by the fluid molecules in the direction parallel to the interface at time $t$. As shown in Fig.~\ref{fig:fig3}(c), $\lambda$ displays a non-monotonous behavior as the inverse screening length $\ell_{\rm TF}^{-1}$ increases from insulating to metallic conditions. Interestingly, the friction $\lambda$ reaches a maximum at a screening length $\ell_{\rm TF}^{-1} \sim 2.7~$nm$^{-1}$ where a maximum in the conductivity $\sigma$ of the Virtual Thomas-Fermi fluid is observed [Fig.~\ref{fig:fig1}(c)]. This result suggests a direct correlation between the electrostatic screening in the metal and the dynamic coupling and underlying interfacial transport at the fluid/solid interface. In fact, as shown in the inset of Fig.~\ref{fig:fig3}(a), $\lambda$ is found to be directly proportional to the sum $\Phi_{\rm{ws}}$ over every relaxation mode $(k,\omega)$ of the water surface response $\operatorname{Im}[g_w(k,\omega)]$ multiplied by its counterpart for the solid $\operatorname{Im}[g_s(k,\omega)]$:
\begin{equation}
	\Phi = \sum_{k,\omega} \operatorname{Im}[g_w(k,\omega)] \operatorname{Im}[g_s(k,\omega)]
	\label{eqmodecoupling}
\end{equation}
The fact that our Virtual Thomas-Fermi approach verifies the latter expression is an expected result. It reflects that the friction parameter is given by the overlap of the charge dynamic structure factors for the solid and liquid. As proposed in Ref. \cite{kavokine2022fluctuation}, such an expression can be derived from a theoretical treatment based on quantum mechanics by considering the product of the uncoupled solid and water surface responses multiplied by a prefactor that describes the solid/fluid coupling ($k,\omega$) for each mode. In contrast, by considering directly the coupled surface responses for the water film and the Virtual Thomas-Fermi fluid, we do not need here to account for such prefactor so that the friction parameter $\lambda$ is directly given by Eq. (\ref{eqmodecoupling}). To illustrate how the overlap of the surface responses govern the friction parameter, Fig.~\ref{fig:fig3}(b) shows the surface responses of the solid, ${\rm Im} [g_s(k,\omega)]$ (contour plot), and for the water film, ${\rm Im} [g_{\rm w}(k,\omega)]$ (white line). The data are provided for the 4 representative inverse screening lengths $\ell_{\rm TF}^{-1}$ which are highlighted in Fig.\ref{fig:fig3}(a). As expected from the formal treatment above, the friction parameter is found to be maximum for $\ell_{\rm TF}^{-1} \sim 2.7$~$\mathrm{nm}^{-1}$ where the overlap between the water and solid surface responses is maximum. In fact, analysis of the surface responses as a function of the screening length indicates that the water surface response is only weakly affected by $\ell_{\rm TF}$ while that for the the Virtual Thomas-Fermi increases towards higher energies $\hbar \omega$ upon decreasing $\ell_{\rm TF}$. As a result, the overlap between the water and solid surface responses first increases with $\ell_{\rm TF}$ to reach a maximum and then decreases upon further increasing $\ell_{\rm TF}$. Qualitatively, recalling that the system employed here does not involve any underlying atomic structure, the results above underscore that friction at the fluid/solid interface is not of static nature only. In contrast to the conventional picture, the friction as observed in the present work is a dynamical phenomenon whose microscopic origin lies in the electrostatic relaxations within the metallic surface and their coupling with the molecular dynamics of the interfacial fluid. 

\noindent {\bf Non-equilibrium transport and electrohydrodynamic drag}\\
\noindent Beyond \textit{quantum friction} and the associated energy dissipation, the dynamic coupling at the solid/fluid interface leads to novel out-of-equilibrium processes. Notably, through this coupling, a molecular  flow in the fluid phase is expected to generate an electrical current in metals~\cite{coquinot2024hydroelectric}. Within the framework of the linear response theory and Onsager's reciprocity rule, this so-called electrohydrodynamic drag mirrors a reciprocal effect called \textit{quantum osmosis} where a charge current in the solid induces fluid flow at their interface. Following the work by Coquinot et al.~\cite{coquinot2024hydroelectric}, the force acting on charge carriers upon applying an electrostatic potential difference $\Delta V$ to the solid along the $x$-direction relates to the pressure difference $\Delta P$ in the fluid through the following Onsager matrix:
\begin{equation}
    \begin{bmatrix}
        Q \\
        I
    \end{bmatrix}
    =
    \begin{bmatrix}
        M_{\rm hh} & M_{\rm eh} \\
        M_{\rm eh} & M_{\rm ee}
    \end{bmatrix}
    \begin{bmatrix}
        \Delta P \\
        \Delta V
    \end{bmatrix},
    \label{eq:onsager}
\end{equation}
In this matrix, the flow rate $Q = S_{\perp}^{\rm w} v_{\rm w}$ is defined as the water velocity $v_{\rm w}$ multiplied by the cross-section area $S_{\perp}^{\rm w} = d_{\rm w}l_y$ perpendicular to the water flow. Similarly, the charge current $I=\rho_-q_{\rm TF} S_{\perp}^{-} v_{-}$ is defined as the density of charge carriers  $q_{\rm TF} \rho_-$) multiplied by the cross-section area $S_{\perp}^{\rm w}$ and the charge velocity $v_{-}$.
As discussed below, $M_{\rm ee}$ relates to the solid conductivity $\sigma$, $M_{\rm hh}$ is linked to the fluid permeability, and $M_{\rm eh}$ is the electrohydrodynamic drag coefficient.

\noindent The electrohydrodynamic drag coefficient $M_{\rm eh}$, which links the molecular flow rate to the driving force $\nabla V = -\Delta V / l_x$, can be determined as follows. First, we write the equation of motion (force balance) for the liquid in the steady state: $\Delta P d_{\rm w} L_y = L_xL_y \left[ \lambda_{\rm w}^0v_{\rm w} + \lambda_{\rm eh}(v_{\rm w} - v_{-}) \right]$. Second, we write the equivalent equation of motion for the solid phase in the steady state: $\rho_-q_{\rm TF} \Delta V d_{\rm TF} l_y = l_xl_y\left[ \lambda_{\rm e}^0 v_- + \lambda_{\rm eh}(v_{\rm e} - v_{\rm w})\right]$. By combining these two equations, we arrive at:
\begin{equation}
    M_{\rm eh} = \frac{\rho_- q_{\rm TF} d_{\rm TF} d_{\rm w} l_y}{l_x} \frac{\lambda_{\rm eh}}{\lambda_{\rm w}\lambda_{\rm e} - \lambda_{\rm eh}^2},
    \label{eq:Mhe} 
\end{equation}
with
\begin{equation}
    \lambda_{\rm w} = \lambda_{\rm w}^0 + \lambda_{\rm eh}, \quad \lambda_{\rm e} = \lambda_{\rm e}^0 + \lambda_{\rm eh}.
    \label{eq:friction}
\end{equation}
In Eq.~(\ref{eq:friction}), the first equality states that the interfacial friction $\lambda_{\rm w}$ involved in the water flow decomposes into two contributions: (1) an intrinsic friction $\lambda_{\rm w}^0$ driven by the static and structural properties of the solid surface and (2) the quantum friction $\lambda_{\rm eh}$ corresponding to the dynamic contribution arising from the coupling of the charge dynamic spectra of the solid and fluid. Analogously, the second equality states that the interfacial friction $\lambda_{\rm e}$ that resists to the charge current within the metal also decomposes into two contributions: (1) an intrinsic contribution $\lambda_{\rm e}^0$ related to the scattering of the negative charges within the metal and (2) the same quantum friction $\lambda_{\rm eh}$ which is involved in both the electrical current and the water molecular flow. In the limit $\lambda_{\rm eh} \to 0$, it follows from Eq.~(\ref{eq:Mhe}) that $M_{\rm eh} \to 0$, so that no electrohydrodynamic drag should be observed as expected for an insulating material. Conversely, as $\lambda_{\rm eh} \to \infty$, $M_{\rm eh} \propto 1/(\lambda_{\rm w}^0 + \lambda_{\rm e}^0)$. The analysis above indicates that maximizing electrohydrodynamic drag can be achieved through large quantum friction $\lambda_{\rm eh}$. In that case, the dynamic coupling at the solid/fluid interface can be maximized by minimizing the solid structure-induced friction in the fluid flow $\lambda_{\rm w}^0$ and/or the electrostatic friction within the solid $\lambda_{\rm e}^0$. This important insight paves the way for the design of novel devices where such friction contributions are tuned and harnessed to control and stimulate molecular and charge flows at the nanoscale. In particular, the above equations suggest that systems with low intrinsic frictions -- such as those involving superfluids or superconductors -- are promising candidates to achieve resistance-free flow and optimal electrohydrodynamic drag.

\noindent Beyond the formal analysis above, the methodology based on Virtual Thomas-Fermi fluids appears as ideal to investigate the electrohydrodynamic drag. Using this strategy in which the charge conductivity within the metal is explicitly taken into account (albeit in an effective fashion), we consider the following simulation set-up to probe electrohydrodynamic drag effects at a fluid/metal interface. Let us consider a metallic material described using the Virtual Thomas-Fermi strategy in which the cations are frozen so that the electrical conductivity corresponds to the transport of negative charge carriers. As a practical situation, we build a simulation box in which a water thin film is sandwiched between two Virtual Thomas-Fermi fluids. As shown in the Supplementary Material, this set-up modification leads to an increase in the liquid/solid friction coefficient as the solid surface responses are  modified due to the frozen cations. Yet, despite this quantitative difference, all the results obtained for this modified set-up remain qualitatively equivalent to those reported in the first part of the present paper. In particular, as detailed in the Supplementary Material, the maximum friction occurs at the same screening length of approximately $2.7~\mathrm{nm}^{-1}$.  

\begin{figure}[h!]
	\centering
	\includegraphics[width=0.7\textwidth]{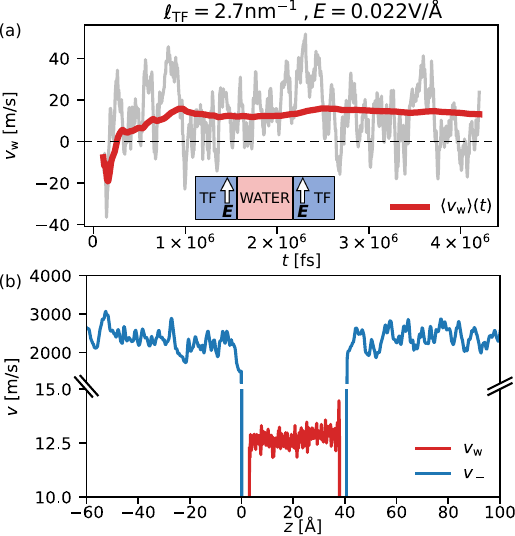}
	\caption{{\bf Non-equilibrium flow as induced through electrohydrodynamic drag.}
		Non-equilibrium flows measured for $\ell_{\rm TF}^{-1} = 2.7~\mathrm{nm}^{-1}$:  (a) Water velocity profile as a function of time along with its time window-averaged value (red line). Despite large statistical fluctuations, the averaged velocity converges to a steady state for times  $t \sim 10^6~\mathrm{fs}$. The inset represents a sketch of the system, where an electrostatic field $E$ is applied to the mobile negative charges in the solid to generate a charge current.
		(b) Solid and liquid velocity profiles across the nanochannel with $z$ representing the direction perpendicular to the solid/fluid interface. The inset illustrates the simulation setup, where the system is driven out of equilibrium by applying an electrostatic field to the negative charges, while the cations are immobile. The dynamic coupling observed in the spectra from Fig.~\ref{fig:fig2} and Fig.~\ref{fig:fig3} leads to an electric field-induced water flow.}
	\label{fig:fig4}
\end{figure}

\noindent To investigate non-equilibrium transport of a water film at a metallic interface, we selected a screening length $\ell_{\rm TF}^{-1} = 2.7~\mathrm{nm}^{-1}$. This value was selected as it maximizes the fluid/solid friction and, thus, enhances any dynamic coupling at this interface. As illustrated in the inset of Fig.~\ref{fig:fig4}(a), an electrostatic field $E$ parallel to the fluid/metal interface is applied to the metallic charge carriers at a time $t = 0$. Fig.~\ref{fig:fig4}(a) shows the average molecular velocity $v_w(t)$ in the water film as a function of time $t$ (we also show in the figure a red line which indicates the time window-averaged value of the data). Despite the expected large fluctuations for such nanoscale films, the velocity stabilizes around $v_w \sim 13~$m/s. This confirms that the applied electrical field in the solid induces a water flow at the interface through quantum friction. Fig.~\ref{fig:fig4}(b) presents the steady-state velocity profiles of water molecules $v_w(z)$ and the metallic ions $v_-(z)$ across the interface. On the one hand, as expected, the electrostatic field $E_x$ induces a charge flow whose velocity is given by $v_- \sim \sigma E_x$.  On the other hand, despite the absence of any pressure gradient (or any other thermodynamic gradient applied to the fluid molecules), the electrostatic field also induces a marked water motion through the so-called electrohydrodynamic drag. While previous works have predicted such effects using formal quantum mechanics treatments,\cite{kavokine2022fluctuation} the results in Fig.~\ref{fig:fig4} provide the first direct account of this intriguing Onsager effect. We stress that this phenomenon cannot be explained without considering both interfacial friction and the conductivity of the charge carriers (in fact, for a non-conductive solid, no continuous electrical current would be observed in any case). 

\begin{figure}[h!]
	\centering
	\includegraphics[width=0.8\textwidth]{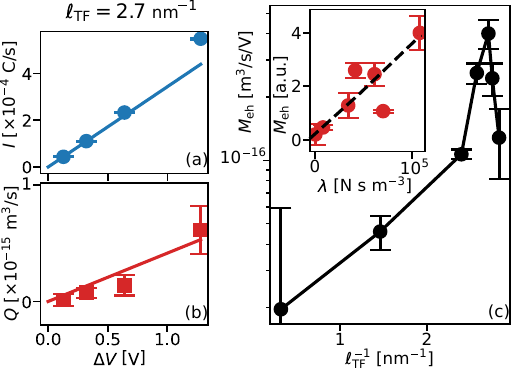}
	\caption{{\bf Linear response theory and electrohydrodynamic drag coefficient.}
    (a-b) Linear response regime for $\ell_{\rm TF}^{-1} = 2.71~\mathrm{nm}^{-1}$ showing the response of ions (a) and water (b) to an applied electrostatic potential difference $E \sim \Delta V$. The solid line represents a linear fit where the slope is related to the electrical conductivity of the solid in (a) and the electrohydrodynamic drag in (b).
    (c) Onsager transport coefficient corresponding to the electrohydrodynamic drag $M_{\rm eh}$ as inferred from the following linear response relation $Q = M_{\rm eh}\Delta V$. As shown in the inset, the maximum $M_{\rm eh}$ is observed at a screening length $\ell_{\rm TF}$ which maximizes the friction parameter $\lambda$ (the latter, which was discussed in a previous figure, was determined from equilibrium molecular dynamics simulations).
	}
	\label{fig:fig5}
\end{figure}

\noindent To shed more light on the electrohydrodynamic drag, Fig.~\ref{fig:fig5}(a-b) shows the system's response under varying potential differences $\nabla V$. As expected, for low to intermediate electrostatic potential differences, both the electrical current within the solid and the molecular flow within the interfacial fluid are proportional to the driving force inducing transport. It was verified that the slope in Fig.~\ref{fig:fig5}(a) -- which corresponds to $M_{\rm ee}$ in Eq.~(\ref{eq:onsager}) -- matches the solid's intrinsic conductivity $\sigma$ as determined from equilibrium molecular dynamics ($M_{\rm ee} = d_{\rm TF} l_y \sigma/l_x$, see Supplementary Material). Fig.~\ref{fig:fig5}(b) presents the induced water flow rate $Q$, which is driven indirectly by the applied potential difference $\Delta V$. As predicted within the linear response framework in Eq.~(\ref{eq:onsager}), $Q$ scales with $\Delta V$ with a proportionality factor corresponding to the electrohydrodynamic drag coefficient $M_{\rm eh}$. Upon examining $M_{\rm eh}$ as a function of $\ell_{\rm TF}^{-1}$, we observe a clear optimum. Notably, this optimum in $M_{\rm eh}$ is observed at a screening length where a maximum is also observed for the friction coefficient $\lambda = \lambda_{\rm h}$ [see inset of Fig.~\ref{fig:fig5}(b)]. Eq.~(\ref{eq:Mhe}) allows us to rationalize as follows this intimate relation between these two transport coefficients. When the screening length is such that the friction coefficient vanishes ($\lambda_{\rm eh} \to 0$), the electrohydrodynamic drag disappears so that $M_{\rm eh} = 0$. Conversely, when the fluid/solid friction reaches its maximum at $\ell_{\rm TF}^{-1} \sim 2.7~\text{nm}^{-1}$, $M_{\rm eh}$ exhibits a pronounced maximum. In the latter case, the ratio in Eq.~(\ref{eq:Mhe}) overall increases when increasing the liquid/solid friction through $\lambda_{\rm h}$. Indeed, in our simulations, intrinsic friction terms -- those unrelated to the coupling between the fluid and solid -- stem from distinct sources: atomic interactions with surface roughness for water (minimal here due to the reflective wall) and electrofriction with frozen cations for anions (the latter acts as an effective viscosity in the Virtual Thomas-Fermi fluid). The strong correlation between the electrohydrodynamic drag coefficient $M_{\rm eh}$ and the friction coefficient $\lambda = \lambda_{\rm h}$ in Fig.~\ref{fig:fig5}(c) confirms  this interpretation. Notably, the peak in $M_{\rm eh}$ coincides with the friction maximum, underscoring friction's dual role both as a resistance parameter and as a mechanism through which the electrohydrodynamic drag occurs.

The results above highlight that optimal electrohydrodynamic drag emerges from a fine balance between the different friction contributions -- namely, the intrinsic friction terms $\lambda^0$ and the cross-coupling term $\lambda_{\rm eh}$. A too weak friction leads to minimal coupling, while a too strong friction makes the flow vanishingly small. In this context, our findings provide novel  insights into the fundamental mechanisms governing interfacial transport in hybrid fluid/metal systems with potential implications for energy harvesting in nanofluidic and electrokinetic devices.


	\noindent \textbf{Discussion} \\
	\noindent We investigated the physical coupling at fluid/metal interfaces which arises from the impact of electrostatic screening and charge relaxation dynamics in the metal on the vicinal fluid. Using an effective molecular approach that captures in a simple but realistic fashion the relaxation modes and conductivity of charge carriers within the metal, we found that the dynamic electrostatic friction induces a marked slowdown in the relaxation of the charge carriers and fluid molecules when the fluid and metal are set in contact. We identified that the overlap between the fluid and solid charge dynamic structure factors is the microscopic ingredient that governs the interfacial friction as involved in nanofluidic transport. As a result, the friction is found to vary in a non-monotonous fashion upon increasing the surface metallicity with a maximum as the solid excitation spectrum strongly overlaps with the charge dynamic structure factor of the fluid phase. Of particular relevance to both fundamental and practical aspects, we also reported a direct observation of the electrohydrodynamic drag through which a water flow is induced by an electrical current within the metallic material. This raises new challenging questions on the complex behavior of fluids in the vicinity or confined within metallic surfaces with important applications in sight such as electrowetting for energy storage, adsorption/catalysis, etc. In particular, by shedding light onto the fundamental mechanisms involved in the blooming field of interfacial transport at metal/fluid interfaces, the present work provides  guidelines for the design of devices with unprecedented performance. 

\begin{methods}
		
\subsection{Molecular Dynamics simulations.}
All simulations were performed using the LAMMPS package \cite{LAMMPS}. The system consists of a solid region and a liquid region. The solid region contains 5000 Thomas-Fermi particles (2500 cations and 2500 anions) with masses $ m_- = 0.005~$g/mol and $m_+ = 0.1~$g/mol, respectively, and a number density $ \rho_{\rm TF} = 0.0263$~\AA{}, chosen to prevent crystallization. The interaction potential between cations and anions includes a purely repulsive potential, described by $U(r) = \frac{E}{r^8}$, where $E = 10^3~$kcal/mol$\cdot$\AA{}$^3$, and electrostatic interactions, which are calculated using the PPPM method with an accuracy of $10^{-4}$ and a cutoff distance of $12.5$~\AA{}.
The liquid region consists of 1000 water molecules modeled using the SPC/E water model \cite{berendsen1987missing}, with a number density $\rho_{\rm w} = 0.033~$\AA{}$^{-3}$. Interactions between particles of different types are governed solely by Coulomb interactions, also calculated with the PPPM method under the same parameters. The lateral dimensions of the simulation box are 
$ L_x = L_y = 29.5~$\AA{}. Reflective walls, positioned $d_{\rm gap}= 3~$\AA{} apart, separate the Thomas-Fermi and water regions. Additional reflective walls were applied along the box boundaries in the $z$-direction (perpendicular to the interface) to prevent particle loss, while periodic boundary conditions were imposed in the $x$- and $y$-directions (parallel to the interface). The timestep is set to $0.1~$fs. Further details on the dependence of friction and conductivity on system parameters are provided in the SM.
Equilibrium simulations consisted of one solid and one liquid region, as depicted in Fig.~\ref{fig:fig1}a. The system was thermostated at $T=300~$K, using a Nos\'e-Hoover thermostat, with separated thermostat applied in the solid and liquid regions, both set to $300~$K. The system was equilibrated for $10^4~\mathrm{fs}$, followed by a production run of $10^5~\mathrm{fs}$, which was long enough for all the time-correlation functions studied in this work to decay to zero.
Non-equilibrium simulations were performed with a setup similar to the equilibrium case, except that the liquid is confined between two solid regions. In this configuration, Thomas-Fermi particles in the right wall do not interact with those in the left wall. Departing from an equilibrated configuration (i.e., the last snapshot of the equilibrium molecular dynamics run), an electrostatic field was applied to the anions at $t=0$~fs, while integrating the water's equations of motion. This was done during an equilibration run of $10^5~$fs, allowing the Thomas-Fermi fluid to reach a steady state with constant velocity. Once steady-state conditions were achieved, data was collected in a production run lasting over $4\cdot 10^6$~fs. 
To address non-physical temperature profiles observed with Nos\'e-Hoover thermostating in non-equilibrium conditions (see SM), we employed Langevin thermostating. The $x$-direction, parallel to the walls, was not thermostated, as an external force $F_x$ was applied to the anions in this direction. The cations were frozen in their equilibrium configuration to ensure a net charge flow. As in the equilibrium simulations, separate thermostats were applied to each region of space: one to the left wall, one to the water region, and another to the right wall.
To validate the method, we measured the friction coefficient from equilibrium simulations using Langevin thermostating in the $y$ and $z$ directions and obtained results consistent with those from Nos\'e-Hoover thermostating in the $x$ direction. This agreement implies that the choice of thermostat does not significantly affect the dynamics in the direction of the flow (see SM). Error bars reported throughout this paper were computed within a $95\%$ confidence level. Gaussian filters were applied to smooth the spectra shown in Figs.~\ref{fig:fig2} and \ref{fig:fig3}.

\subsection{Screening length determination.}
Following previous work \cite{schlaich2022electronic}, while Eq.~(\ref{eq:ellTF}) provides an estimate for the order of magnitude of the screening length $\ell_{\rm TF}$, we directly measured it in our simulations using capacitance measurements. To achieve this, we constructed a system consisting of two solid regions composed of Thomas-Fermi particles with identical parameters as in the other simulations, separated by a vacuum gap of $L_{\rm vac} = 30~$\AA{}. The system was sandwiched between two electrodes with surface charge densities of $+\Sigma$ and $-\Sigma$, where $\Sigma = 0.166~$C/m$^2$.
The simulation parameters, including the timestep and thermostating method, were consistent with those used in equilibrium measurements. Nos\'e-Hoover thermostats were employed at $300~$K, with equilibration and production runs lasting $10^5~$fs each. Equivalent results were also obtained using Langevin thermostating. By performing simulations for various values of $q_{\rm TF}$, we determined the effective screening length $\ell_{\rm TF}$ based on capacitance measurements, using the relation $\frac{1}{C} = ( L_{\rm vac} + 2 \ell_{\rm TF})/\varepsilon_0 $. Here, $C$ is the capacitance, and $\varepsilon_0$ is the vacuum permittivity. 
		
\subsection{Conductivity determination.}
The conductivity $\sigma$ reported in Fig.~\ref{fig:fig1}c was determined using two complementary approaches. The first approach uses the electric current autocorrelation function:
\begin{equation}
\sigma = \frac{1}{3 V k_{\rm B} T} \int_0^\infty {\rm d}t \, \langle \vb{J}(0) \cdot \vb{J}(t) \rangle,
\end{equation}
where $V$ is the volume of the solid region, $k_{\rm B}$ is the Boltzmann constant, and $\vb{J}(t) = \sum_{i=1}^N q_i(t) \vb{v}_i(t)$ is the electric current, with $q_i$ being the charge and $\vb{v}_i$ the velocity of the $i$-th Thomas-Fermi particle.
The second approach is based on the collective translational dipole moment of a particular species $k$, where $k = +$ for cations and $k = -$ for anions. The dipole moment is given by $\vb{M}_J^k(t) = \sum_{i \in k} q_i \cdot \vb{r}_i(t)$, with $\vb{r}_i(t)$ being the time-dependent position of the $i$-th ion. The total collective dipole moment is then $\vb{M}_J(t) = \vb{M}_J^+(t) + \vb{M}_J^-(t)$. The conductivity is determined from the mean squared displacement of the collective dipole:
\begin{equation}
\sigma = \frac{\langle \Delta M_J^2 \rangle}{V k_{\rm B} T t}.
\end{equation}
We verified that both methods yielded equivalent results, and the values shown in the figure correspond to the average of both approaches.

\end{methods}
	
	\bibliography{cecilia}
	
	\begin{addendum}
		\item We acknowledge L. Joly, N. Kavokine, P. Gispert and B. Coquinot for fruitful discussions. We also acknowledge funding from the ANR project TAMTAM (ANR-15-CE08-0008-01). This work was granted access to the HPC resources of IDRIS under the allocation 2025-103932 made by GENCI.
		
		\item[Competing Interests] The authors declare that they have no
		competing financial interests.
		
		\item[Author contributions] C.H.\, B.C.\, and L.B.\ conceived the 
		research. C.H.\ carried out the molecular simulations.
		C.H., B.C.\ and L.B.\ analyzed the data. C.H.\ and B.C.\ wrote the 
		paper with inputs from all authors.
		
		\item[Correspondence] Correspondence and requests for materials
		should be addressed to C.H.
	\end{addendum}
	
\end{document}


\title{Supplemental Material: Fluids at an electrostatically active surface:  Optimum ininterfacial friction and electrohydrodynamic drag}

\author{Cecilia Herrero}

\affiliation{\ill}
\affiliation{\ens}

\author{Lyd\'eric Bocquet}

\affiliation{\ens}

\author{Benoit Coasne}
\affiliation{\ill}
\affiliation{\liphy}

\date{\today}

\maketitle

\setcounter{equation}{0}
\setcounter{figure}{0}
\setcounter{table}{0}
\setcounter{page}{1}
\renewcommand{\theequation}{S\arabic{equation}}
\renewcommand{\thefigure}{S\arabic{figure}}
\renewcommand{\bibnumfmt}[1]{[S#1]}
\renewcommand{\citenumfont}[1]{S#1}
\renewcommand\thesubsection{\arabic{subsection}}
\renewcommand\thesubsubsection{\arabic{subsection}.\arabic{subsubsection}  }

\textbf{Conductivity and friction dependence on system parameters.}
\begin{figure}[h!]
    \centering
    \includegraphics{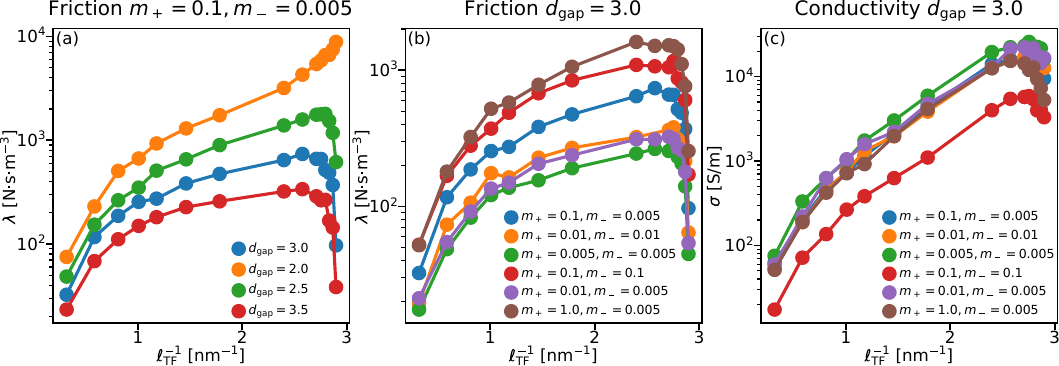}
    \caption{(a) Friction coefficient for different liquid-solid (i.e., water-Thomas Fermi liquid) distances $d_{\rm gap}$, corresponding to the vacuum size between the respective reflective walls. We observe the expected shift in the friction's order of magnitude, decreasing with increasing $d_{\rm gap}$. However, the overall behavior, such as the maximum caused by the optimal dynamic overlap in the liquid-solid spectra (discussed in the main text), remains consistent across all ranges except for the smallest $d_{\rm gap} = 2.0~\text{\AA}$, where the system becomes unstable due to the proximity of the two regions.
    (b) Friction coefficient for different anion-cation masses. The mass variation affects the order of magnitude of the solid's dynamic spectra but does not alter the overall behavior described in the text. 
    (c) Conductivity measurements for varying anion-cation masses. The conductivity is largely unaffected by the mass, except for the extreme case $m_+ = m_- = 0.1~\text{g/mole}$. Note that $d_{\rm gap}$ is given in $\text{\AA}$, and $m$ is in $\text{g/mole}$.}
    \label{fig:fig_s2}
\end{figure}

\newpage 

\textbf{Friction results for Thomas-Fermi model with frozen cations.}
\begin{figure}[h!]
    \centering
    \includegraphics[width=0.7\textwidth]{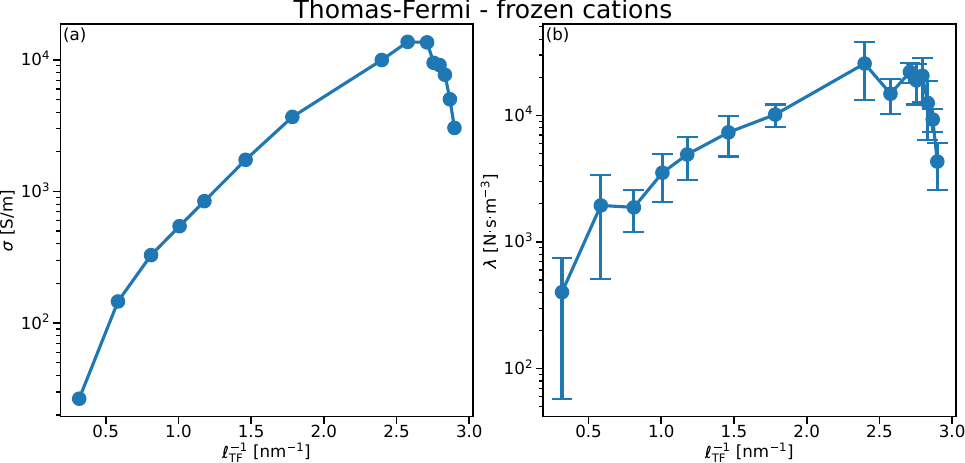}
    \caption{(a) Conductivity and (b) friction coefficient, calculated using the Green-Kubo integral, for the virtual Thomas-Fermi fluid model with frozen cations. The results show an increase of approximately one order of magnitude in the absolute value of friction compared to the fully thermostated system presented in the main text. Despite this difference, key features of the system's behavior are preserved, including the maximum in friction observed at $\ell_{\rm TF} \approx 2.7~\text{nm}^{-1}$. These results align with the expected dynamics for this simplified model.}
    \label{fig:fig_s1}
\end{figure}

\textbf{Discussion on the thermostat choice for non-equilibrium simulations.}
\begin{figure}[h!]
    \centering
    \includegraphics[width=0.7\textwidth]{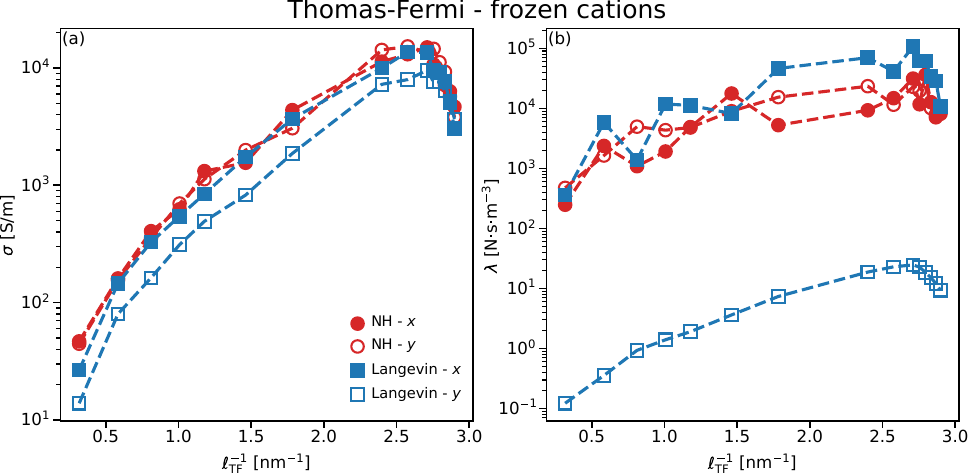}
    \caption{(a) Conductivity and (b) friction coefficient, determined from equilibrium measurements, for both Nos\'e-Hoover (NH) and Langevin thermostats applied in the $y$ and $z$ directions. The results demonstrate that, in the $x$ direction, the Langevin thermostat does not significantly alter the system's transport coefficients compared to the NH results. This supports the choice of using a Langevin thermostat for the non-equilibrium measurements discussed in the main text, as the liquid becomes otherwise too hot under the NH thermostat for the non-equilibrium simulations.}
    \label{fig:fig_s3}
\end{figure}

\textbf{Verification linear response and conductivity agreement}
\begin{figure}[h!]
    \centering
    \includegraphics[width=0.5\textwidth]{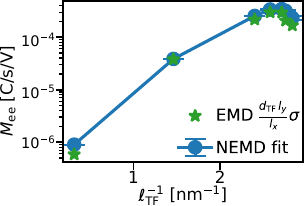}
    \caption{Onsager response coefficient describing transport in the Thomas-Fermi fluid, where the solid current defines $M_{\rm ee}$ as $I=M_{\rm ee}\Delta V$. We verify that the maxima in $M_{\rm ee}$ correspond to peak in conductivity, as determined from equilibrium molecular dynamics (EMD) simulations.}
    \label{fig:fig_s4}
\end{figure}
